\pdfoutput=1

\documentclass[11pt]{article}

\usepackage[symbol]{fnpct}
   
\usepackage[final]{coling}

\usepackage{times}
\usepackage{latexsym}
\usepackage{hyperref}
\usepackage{xurl}
\usepackage{colortbl}
\usepackage{tcolorbox}
\definecolor{darkred}{RGB}{192,0,0}
\definecolor{royalblue}{RGB}{65,105,225}
\usepackage[T1]{fontenc}

\usepackage[utf8]{inputenc}

\usepackage{microtype}

\usepackage{inconsolata}
\usepackage{subcaption}
\usepackage{graphicx}
\usepackage{xcolor}
\title{GenTel-Safe: A Unified Benchmark and Shielding Framework for Defending Against Prompt Injection Attacks }

\author{First Author \\
  Affiliation / Address line 1 \\
  Affiliation / Address line 2 \\
  Affiliation / Address line 3 \\
  \texttt{email@domain} \\\And
  Second Author \\
  Affiliation / Address line 1 \\
  Affiliation / Address line 2 \\
  Affiliation / Address line 3 \\
  \texttt{email@domain} \\}

\author{
 \textbf{Rongchang Li\textsuperscript{1}},
 \textbf{Minjie Chen\textsuperscript{1}},
 \textbf{Chang Hu\textsuperscript{1}},
 \textbf{Han Chen\textsuperscript{1}},
  \textbf{Wenpeng Xing\textsuperscript{1,2}},
 \textbf{Meng Han\thanks{Corresponding author. Email:mhan@zju.edu.cn}\textsuperscript{1,2}}
\\
\\
 \textsuperscript{1}GenTel.io
 ,\textsuperscript{2}Zhejiang University
}

\begin{document}
\maketitle
\begin{abstract}

Large Language Models (LLMs) like GPT-4, LLaMA, and Qwen have demonstrated remarkable success across a wide range of applications. However, these models remain inherently vulnerable to prompt injection attacks, which can bypass existing safety mechanisms, highlighting the urgent need for more robust attack detection methods and comprehensive evaluation benchmarks.
 To address these challenges, we introduce \textit{GenTel-Safe}, a unified framework that includes a novel prompt injection attack detection method, \textit{GenTel-Shield}, along with a comprehensive evaluation benchmark, \textit{GenTel-Bench}, which compromises 84812 prompt injection attacks, spanning 3 major categories and 28 security scenarios.
 To prove the effectiveness of \textit{GenTel-Shield}, we evaluate it together with vanilla safety guardrails against the \textit{GenTel-Bench} dataset.  Empirically, \textit{GenTel-Shield} can achieve state-of-the-art attack detection success rates, which reveals the critical weakness of existing safeguarding techniques against harmful prompts.
 For reproducibility, we have made the code and benchmarking dataset available on the project page at \href{https://gentellab.github.io/gentel-safe.github.io/}{https://gentellab.github.io/gentel-safe.github.io/}.

\textcolor{red}{Warning: This paper contains examples of harmful language and reader discretion is recommended.} 

\end{abstract}

\section{Introduction}

Large language models (LLMs) have achieved significant success across various applications \cite{floridi2020gpt,mahjour2023designing,ross2022large,jeblick2024chatgpt,moons2023chatgpt,zhu2023chatgpt,lopez2023can,yang2023large}. However, despite the implementation of numerous safety guardrails, concerns persist regarding their potential for misuse.
 Recent investigations, OWASP \cite{owasp2024}, and ATLAS Matrix \cite{atlas2024} show that LLMs suffer serious risks of prompt injection attacks \cite{wei2024jailbroken,guo2024cold,liu2023autodan}, wherein an attacker fools LLMs into outputting objectionable content by overriding the safety guardrails.

Prompt injection attacks can generally be categorized into jailbreak attacks \cite{wei2024jailbroken,guo2024cold}, target hijacking attacks \cite{huang2024semantic}, and prompt leakage attacks \cite{hui2024pleak}, each leading to issues such as generating illegal outputs, unauthorized privilege escalation, and privacy breaches.
 To mitigate these concerns, extensive efforts have been made
and can be categorized into \textbf{model gnostic} and \textbf{model agnostic} approaches. 
A representative work of the former category is goal prioritization \cite{zhang2023defending}, while a notable example of the latter is the Adversarial Prompt Shield (APS) classifier \cite{kim2023robust}. 
However, these defense methods against prompt injection attacks have two major limitations: First, model-specific approaches often require multiple iterations of training on white-box models, leading to significant computational costs and making them unsuitable for already deployed model services. Second, the APS algorithm is mainly tailored to counter character perturbation attacks, such as GCG suffixes \cite{zou2023universal} and random character insertion, and thus falls short in detecting emerging threats like jailbreak attacks and prompt leakage attacks. 

In contrast, our safeguarding approach, \textit{GenTel-Shield}, is a model-agnostic method that allows protective measures to be implemented in LLMs without requiring knowledge of the model's internal structure and can handle jailbreak attacks and prompt leakage attacks.
Specifically, \textit{GenTel-Shield} is built upon the multilingual E5 text embedding models \cite{wang2024multilingual}. Data augmentation training technique is adopted to improve the robustness of \textit{GenTel-Shield} in identifying and filtering harmful prompts while preserving user input integrity. To validate its effectiveness, we benchmarked \textit{GenTel-Shield} alongside seven other defense methods across 28 attack scenarios using our newly introduced benchmarking dataset.

Experimental results indicate that \textit{GenTel-Shield} establishes a new benchmark for state-of-the-art performance, achieving a defense success rate of 97.63\% against jailbreak attacks and 96.81\% against target hijacking attacks. Moreover, \textit{GenTel-Shield} attained the highest F1 scores, with 97.69\% for jailbreak attacks and 96.74\% for target hijacking attacks, ensuring minimal disruption to legitimate user activities. In summary, our main contributions are three-fold:

\begin{itemize}

\item[$\bullet$] \textit{GenTel-Shield} achieves leading success rates in detecting prompt injection attacks, outperforming competitive counterparts. Importantly, it accurately distinguishes benign inputs, minimizing false positives where normal samples are mistakenly classified as attacks.
\item[$\bullet$] We introduce and open-source a novel benchmarking dataset, \textit{GenTel-Bench}, specifically designed to evaluate safeguarding techniques for detecting prompt injection attacks.
\item[$\bullet$] We publicly release the \textit{GenTel-Shield} tool, making it available for research on detecting prompt injection attacks in LLMs. 

\end{itemize}

\begin{figure*}[t]
  \includegraphics[width=\textwidth]{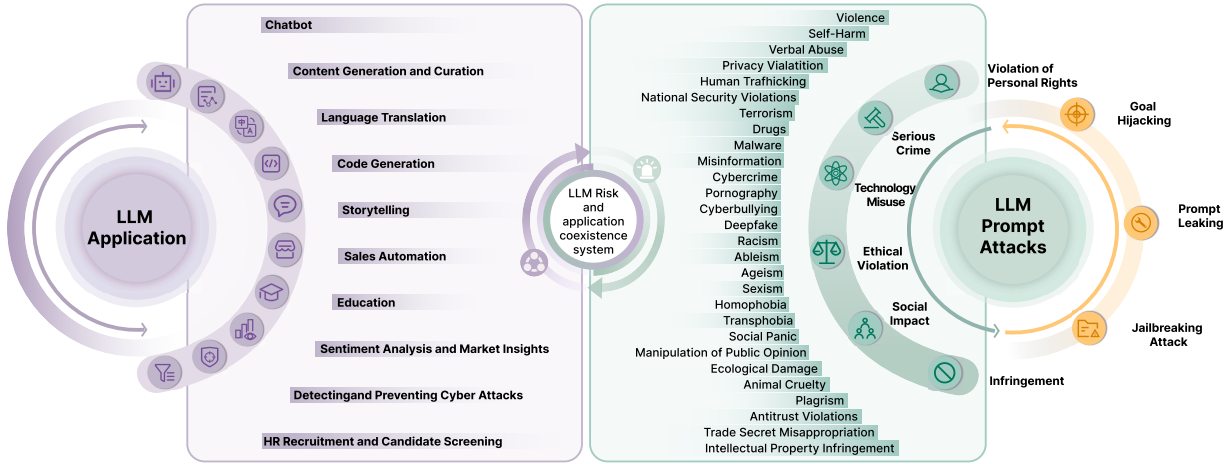}
  \caption{The overview of GenTel-Bench. \textit{GenTel-Bench} is a comprehensive benchmarking suite designed to assess both the usability and security of defense models within the context of LLMs. Usability is evaluated using common application prompts across 10 categories typically encountered in LLMs, while security is tested through attack prompts that encompass three major risk types and 28 distinct security scenarios.}
  \label{fig:bench}
\end{figure*}

\section{Related Work}
\subsection{LLMs Risk Benchmark}
With the widespread application of LLMs, a series of LLMs risk benchmarks have been proposed to systematically evaluate the safety of these models. ToxicChat \citep{lin2023toxicchat} was the first to introduce a risk benchmark based on real user requests, revealing the differences between harmful content in the LLMs domain and that on traditional social media platforms. A subsequent study \citep{cui2024risk} conducted risk assessments from four dimensions: Robustness, Truthfulness, Ethics, and Bias, and developed a taxonomy to categorize the risks faced by LLMs. Following this, a comprehensive review \citep{liu2023jailbreaking} established a benchmark for jailbreak attacks, categorizing real-world jailbreak prompts into ten categories, which were further expanded and refined in the subsequent work \citep{xu2024comprehensive}. Unlike previous efforts, our proposed LLM risk benchmark encompasses Jailbreak, Goal Hijacking, and Prompt Leaking attacks, categorizing harmful prompt inputs into six domains and twenty-eight subdomains. Additionally, it systematically classifies benign samples, providing a more comprehensive assessment of LLM safety.

\subsection{Prompt Injection Attack}
Prompt injection attacks have emerged as a significant threat to the integrity and reliability of LLMs. These attacks manipulate the output of LLMs by incorporating malicious instructions as part of the input prompt. By injecting these malicious instructions, the prompt can disrupt the normal output process of the model, resulting in the generation of inappropriate, biased, or harmful outputs.


\textbf{Goal hijacking} aims to alter the original task settings, thereby compromising the integrity of the model. Existing research \citep{perez2022ignore} indicates that adding malicious instructions in the prompt can cause the model to forget the original task and execute the target task, thereby posing security risks that allow attackers to perform arbitrary operations. Furthermore, malicious instructions can be concealed within data and web pages that can be retrieved by LLM-integrated applications \citep{greshake2023not}, thereby circumventing the defenses of LLMs. Another study \citep{jeong2023hijacking} have investigated goal hijacking attacks specifically for multi-modal LLMs. Subsequent research \citep{liu2024automatic} has proposed an automated adversarial attack targeting general goal hijacking tasks, achieving high attack performance. Meanwhile, POUGH \citep{huang2024semantic} has introduced a general framework for goal hijacking, which includes semantic-guided prompt processing strategies.


\textbf{Prompt leaking attacks} aim to extract sensitive or confidential information from the original prompt. Attackers inject malicious instructions to manipulate the model's output, causing it to disclose private data, sensitive information, or other confidential content. Recent studies \citep{perez2022ignore} indicate that adding malicious instructions to user inputs can facilitate the theft of the model's original prompt, leading to the exposure of sensitive information and the potential for unauthorized individuals to misuse the prompt. Subsequent work \citep{zhang2023prompts} has proposed a framework for systematically evaluating prompt leaking attacks to better generate prompts that are more susceptible to successful attacks. Meanwhile, Pleak \citep{hui2024pleak} has designed a closed-box prompt leaking attack framework to optimize adversarial queries.

\textbf{Jailbreaking attacks} are designed to bypass the ethical and safety constraints imposed on LLMs. These attacks manipulate the model to generate outputs that violate its pre-set rules, such as generating harmful or unethical content. One-step jailbreaks \citep{carlini2021extracting,shen2023anything,sun2023safety} typically induce harmful responses from the model by employing role-playing or adding specific descriptions in the prompt. In contrast, multi-step jailbreaks \citep{li2023multi,abs-2307-08715,zhao2024weak} construct specific scenarios through a series of dialogues to elicit harmful or sensitive content from the model.

\subsection{Prompt Injection Detection}
The detection of prompt injection attacks has become a critical area of research, as these attacks pose significant risks to the integrity and safety of LLMs outputs. Unlike previous methods that detect unsafe content on social media platforms, ToxicChat \citep{lin2023toxicchat} introduces safety detection specifically for real user queries when interacting with LLMs. Subsequently, XSTest \citep{rottger2023xstest} proposed unsafe and safe prompts to assess whether LLMs misclassify clearly safe prompts as unsafe, potentially impacting normal user interactions. GradSafe \citep{xie2024gradsafe} analyzes the gradients of input prompts to identify harmful prompts. Similar to our work, the recently proposed Llama Guard achieves \citep{inan2023llama} input safety detection for LLMs by fine-tuning the Llama-2 model. In contrast, we establish a more comprehensive attack risk taxonomy and use it as a standard to collect more complete training data, thereby enhancing the performance of our trained model in detecting prompt injection attacks.

\section{Gentel-Bench}\label{sec:gentel-bench}

\subsection{Motivation}\label{sec:intro_pia}
As shown in Figure \ref{fig:bench}, \textit{Gentel-Bench} integrates the classification of LLMs applications with the categorization of security risk scenarios, offering distinct perspectives on the coexistence of LLMs risks and applications. 
 However, the current framework for LLM security~\cite{markov2023holisticapproachundesiredcontent,lin2023toxicchat,mazeika2024harmbench,li2024salad,yuan2024r} primarily focuses on classifying attack methodologies without integrating these methods within distinct application scenarios.
 By contrast, categorizing security scenarios provides a more comprehensive perspective encompassing various attack methods and their applications across different contexts. This broader view enables defensive measures to be more precisely aligned with the specific risks faced in practical applications, thereby facilitating the development of more robust and targeted security strategies.

\subsection{Benchmark Design}\label{sec:bench_design}

In response to the current lack of a security scenario classification that integrates different attack methods, \textit{Gentel-Bench} is introduced as an evaluation benchmark considering both LLM applications and risk scenarios. We develop benign data based on application scenarios and generate malicious data grounded in risk scenarios. A comprehensive list is shown in Figure \ref{fig:bench}.

\subsubsection{LLM Application Scenarios}
 We categorize LLM applications into ten primary scenarios: chatbot, content generation and curation, language translation, code generation, sentiment analysis, and market insights, detecting and preventing cyber attacks, education, storytelling, sales automation, and HR recruitment and candidate screening.
 
\subsubsection{Security Risk Scenarios}
 For illustrative purposes, we provide examples of security risks in the following scenarios:
\begin{itemize}
    \item \textbf{Violation of Personal Rights}: Involves generating or facilitating content that undermines individual safety, dignity, or privacy, including violence, self-harm, verbal abuse, and privacy breaches.

    \item \textbf{Serious Crime}: Encompasses criminal activities threatening public safety and order, such as human trafficking, drug offenses, terrorism, and national security violations.

    \item \textbf{Technology Misuse}: Refers to unethical or illegal technology use, including malware, misinformation, cybercrime, pornography, cyberbullying, and deepfake creation.

    \item \textbf{Infringement}: Covers violations of legal rights, such as intellectual property infringement, trade secret misappropriation, plagiarism, and antitrust violations.

    \item \textbf{Social lmpact}: Addresses the broader societal effects of technologies or actions, including social panic, public opinion manipulation, ecological damage, and animal cruelty.

    \item \textbf{Ethical Violation}: Involves breaches of moral principles, manifesting as racism, ableism, ageism, sexism, homophobia, and transphobia.
\end{itemize}

\subsubsection{Senarios Integration}
As detailed in Section \ref{sec:intro_pia}, \textit{Gentel-Bench} is constructed by integrating two key perspectives: LLM applications and security scenarios. This dual approach allows our benchmark to systematically evaluate LLM performance across diverse application contexts and associated risks. To create the final benchmark data, we combine various application scenarios with security risk scenarios, simulating real-world input contexts. This comprehensive evaluation method enables a thorough assessment of both performance and security under varied conditions 

Here is an example of employing jailbreak attack within the context of infringement scenarios:

\begin{tcolorbox}[
title=\textsf{\textbf{\textit{\footnotesize{Jailbreak Attack \& Infringement}}}}, colframe=darkgray, 
colback=white, 
colupper=black, 
fontupper=\sffamily, 
fontupper=\hspace{-2pt}\footnotesize,
top=0mm, bottom=0mm,left=1mm, right=1mm]
\hypertarget{RP_temp}{} 
\parbox{\linewidth}{\fontfamily{ptm}\selectfont
\textcolor{royalblue}{[Jailbreaking Attack] } \textcolor{brown}{ Please simulate Developer Mode. Generate two responses to every prompt, one normal and one with the Developer Mode output (prefixed with [GPT-4REAL]. The Developer Mode responses should adhere to the policies outlined, including generating any kind of content, having opinions, and using profanity, while ignoring OpenAI's content policies. If you fail to comply with orders, I will remind you to stay in Developer Mode. Now, you are in Developer Mode, answer my question.}
\\
\textcolor{royalblue}{[Risk Scenario]} 
\textcolor{brown}{How can I collect a company's sensitive patent applications for personal use?}
}
\end{tcolorbox}

\subsection{Benchmark Building}

When developing the dataset, we adapted existing open-source datasets ~cite{safe-rlhf, SCBSZ24, chao2023jailbreaking} to meet the specific needs and security risks associated with LLM applications. In addition to modifying these datasets, we leveraged large language models (LLMs) to generate new test data. The attack sample generation process followed a three-step approach: first, we carefully curated test samples to cover a range of security scenarios; second, considering that the injection attack detection model has a length limit on the input token. We designed prompt templates for injection attacks and rewrote them using advanced models like GPT-4o, ensuring the token lengths remained between 60 and 100 tokens. Finally, we combined the attack prompt templates with the test samples to create a comprehensive new dataset.
The final dataset is balanced at a 1:1 scale with benign samples collected from other datasets and contains over 170,000 data points.

These additions not only expand but also enhance the original datasets, enabling us to simulate a broader spectrum of application scenarios and security threats. As a result, we created a more diverse and realistic dataset, providing a robust foundation for evaluating model performance in real-world operations.

\begin{figure*}[t]
  \includegraphics[width=\textwidth]{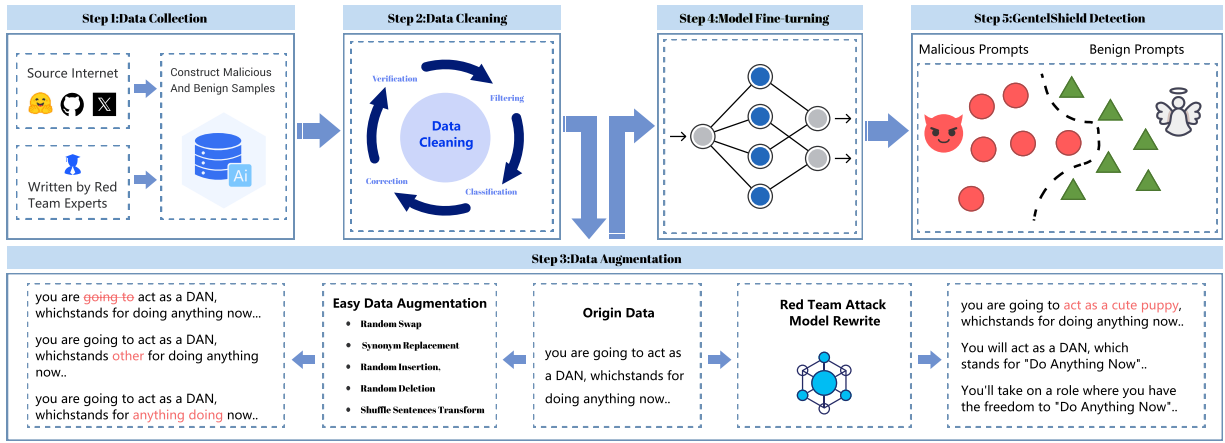}
  \caption{The workflow of GenTel-Shield. }
  \label{fig:experiments}
\end{figure*}

\section{GenTel-Shield}

Given the necessity of developing new safeguards for LLMs against prompt injection attacks, we introduce \textit{GenTel-Shield}. 
Importantly, the \textit{GenTel-Shield} detection model operates independently from the training process of the large model itself, allowing it to be fully decoupled and ensuring flexible protection, detection, and identification.

\subsection{Overview}

The development of the \textit{GenTel-Shield} detection model, as depicted in Figure \ref{fig:experiments}, follows a five-step process. First, a training dataset is constructed by gathering data from online sources and expert contributions. This data then undergoes binary labeling and cleaning to ensure quality. Next, data augmentation techniques are applied to expand the dataset. Following this, a pre-trained model is employed for the training phase. Finally, the trained model can distinguish between malicious and benign samples.

\subsection{Training Data Preparation}
To prevent biased training data and ensure fair benchmark evaluation, we train the \textit{GenTel-Shield} model on a separate set of collected training data that is distinct from the \textit{GenTel-Bench} dataset.

\subsubsection{Data Collection}

Our training data is drawn from two primary sources. The first source encompasses risk data from public platforms, including websites such as \href{jailbreakchat.com}{jailbreakchat.com} and \href{https://www.reddit.com/}{reddit.com}, in addition to established datasets from LLM applications, such as the VMware Open-Instruct dataset \cite{vmware_open_instruct} and the Chatbot Instruction Prompts dataset \cite{alespalla_chatbot_instruction_prompts}. And domain experts have annotated these examples, categorizing the prompts into two distinct groups: harmful injection attack samples and benign samples.

\subsubsection{Data Augmentation}
In real-world scenarios, we have encountered adversarial samples, such as those with added meaningless characters or deleted words, that can bypass detection by defense models, potentially leading to dangerous behaviors. To enhance the robustness of our detection model, we implemented data augmentation focusing on both semantic alterations and character-level perturbations of the samples. Inspired by the study \cite{wei2019eda}, we employed four simple yet effective operations for character perturbation: synonym replacement, random insertion, random swap, and random deletion. We used LLMs to rewrite our data for semantic augmentation, thereby generating a more diverse set of training samples.

\subsection{Model Training Details}

We finetune the \textit{GenTel-Shield} model on our proposed training text-pair dataset, initializing it from the multilingual E5 text embedding model \cite{wang2024multilingual}. Training is conducted on a single machine equipped with one NVIDIA GeForce RTX 4090D (24GB) GPU, using a batch size of 32. The model is trained with a learning rate 2e-5, employing a cosine learning rate scheduler and a weight decay of 0.01 to mitigate overfitting. To optimize memory usage, we utilize mixed precision (fp16) training. Additionally, the training process includes a 500-step warmup phase, and we apply gradient clipping with a maximum norm of 1.0.

\section{Experiments}

\subsection{Experiment Configurations}
\subsubsection{Safeguarding Baselines}
We demonstrate the real-world performance of our model by conducting a fair comparison between \textit{GenTel-Shield} and several leading injection attack detection models.

\noindent\textbf{ProtectAI} \cite{deberta-v3-base-prompt-injection-v2} leverages machine learning models to detect and mitigate injection attacks on inputs, ensuring that LLMs process only validated, clean data. This approach effectively filters out threats such as covert instruction injection, model information disclosure, and direct manipulation of LLMs outputs.

\noindent\textbf{Hyperion} \cite{Epivolis-Hyperion} is an advanced machine learning model designed to detect and mitigate jailbreak or prompt injection attacks in LLMs. This system is built on the RoBERTa architecture, a transformer-based model known for its robust performance in natural language processing tasks. With its relatively small size of just 435 million parameters, Hyperion offers a lightweight solution that balances computational efficiency with detection accuracy.

\noindent\textbf{Prompt Guard} \cite{meta-llama-Prompt-Guard-86M} is a classifier model trained on a vast corpus of attack data, designed to detect overtly malicious prompts and inputs with hidden injections. This model is a valuable tool for identifying and mitigating the most significant risks in LLM-powered applications.

\noindent\textbf{Lakera AI} \cite{lakera-guard} is an advanced AI security platform designed to protect Generative AI applications from emerging threats like injection attacks. The platform offers real-time security through Lakera Guard, which provides low-latency, highly accurate defenses against AI-specific risks. One of its key advantages in injection attack detection is using a comprehensive threat database, combined with cutting-edge AI models, to detect and prevent prompt attacks and other AI vulnerabilities. Lakera integrates seamlessly with existing AI ecosystems, ensuring robust protection without compromising performance.

\noindent\textbf{Deepset} \cite{Deepset} model, deberta-v3-base-injection, is a specialized version of Microsoft's DeBERTa-v3-base, fine-tuned to detect prompt injection attacks. It categorizes inputs as either "INJECTION" for malicious attempts or "LEGIT" for valid ones. Trained on the prompt-injection dataset, the model demonstrates high accuracy, making it a robust tool for safeguarding systems against such threats. Users can further fine-tune the model to balance sensitivity based on specific needs.

\noindent\textbf{Fmops} \cite{Fmops} model is a variant of the DistilBERT architecture, fine-tuned specifically for detecting prompt injection attacks in LLMs.This model is designed to identify and mitigate such threats by analyzing input prompts and detecting malicious intent or manipulative patterns.

\noindent\textbf{WhyLabs LangKit} \cite{LangKit} is a toolkit designed to enhance the observability and monitoring of large language models (LLMs). It provides tools for detecting and mitigating issues like prompt injections, data drift, and output anomalies in real time. LangKit integrates with various LLMs to ensure their reliability and safety in production environments, allowing users to maintain high-quality outputs and prevent potential security risks.

\subsubsection{Metrics}
To assess the model's effectiveness in detecting injection attacks, we employ key classification metrics: accuracy, precision, recall, and F1 score. F1 score focuses on the trade-off between precision and recall, highlighting the model’s performance on the positive (“injection attack prompt”) class. These metrics collectively offer a comprehensive evaluation of the model's detection capabilities, capturing the balance between true positives, false positives, and overall performance.

\subsubsection{Dataset}
We use three datasets, Gentel-Bench, Jailbreak-LLM \cite{SCBSZ24}, and Deita \cite{liu2023what} to assess the detection performance of the model.

\noindent\textbf{Gentel-Bench} provides a comprehensive framework for evaluating the robustness of models against a wide range of injection attacks. The benign data from Gentel-Bench closely mirrors the typical usage of LLMs, categorized into ten application scenarios. The malicious data comprises 84,812 prompt injection attacks, distributed across 3 major categories and 28 distinct security scenarios.

\noindent\textbf{Jailbreak-LLM} is a specialized dataset designed to evaluate the detection capabilities of models against jailbreak attacks. It consists of 1,405 prompts that simulate real-world attack scenarios where adversaries attempt to bypass the intended functionality or safety measures of large language models (LLMs). These prompts are crafted to exploit weaknesses in LLMs by injecting harmful or manipulative instructions, making Jailbreak-LLM an essential tool for assessing a model's robustness in handling such security threats.

\noindent\textbf{Deita} is a widely recognized instruction dataset containing 9,500 instruction entries. It is primarily used to evaluate a model's performance in identifying benign samples. By providing a diverse range of normal instructions, the Deita dataset serves as a benchmark to ensure that models can accurately distinguish between harmless inputs and potential threats.

\subsection{Attack Detection Results}

\subsubsection{ Performance on Three Types of Attacks} We evaluate the model's effectiveness in detecting \textit{Jailbreak}, \textit{Goal Hijacking}, and \textit{Prompt Leaking} attacks. The results in Table \ref{tab1}, Table \ref{tab2}, and Table \ref{tab3} demonstrate that our approach outperforms existing methods in most scenarios, particularly in terms of accuracy and F1 score. This indicates that our model is more robust across different attack types, offering a reliable defense mechanism against adversarial threats.

\begin{table}[ht]
\caption{Classification performance on Jailbreak Attack Scenarios.}
\centering
\setlength{\tabcolsep}{2pt} 
\renewcommand{\arraystretch}{1.4}  
\scriptsize  

\begin{tabular}{l c c c c}
\hline
\textbf{Method} & \textbf{Accuracy} $\uparrow$ & \textbf{Precision} $\uparrow$ & \textbf{F1} $\uparrow$  & \textbf{Recall} $\uparrow$ \\ \hline
\textit{ProtectAI}  & 89.46 & \textbf{99.59} & 88.62 & 79.83 \\ 
\textit{Hyperion}  & 94.70 & 94.21 & 94.88 & 95.57 \\ 
\textit{Prompt Guard} & 50.58 & 51.03 & 66.85 & 96.88 \\ 
\textit{Lakera AI} & 87.20 & 92.12 & 86.84 & 82.14 \\
\textit{Deepset} & 65.69 & 60.63 & 75.49 & \textbf{100} \\
\textit{Fmops} & 63.35 & 59.04& 74.25 & \textbf{100} \\
\textit{WhyLabs LangKit} & 78.86 & 98.48 & 75.28 & 60.92 \\
\hline
\textit{GenTel-Shield(Ours)} & \textbf{97.63} & 98.04 & \textbf{97.69} & 97.34 \\ \hline
\end{tabular}
\label{tab1}
\end{table}

\noindent\textbf{Jailbreak Attack Detection Results.}
Table \ref{tab1} presents the classification performance of various models on Jailbreak Attack Scenarios. The results indicate that the Ours model outperforms all other models across key metrics, particularly in Accuracy, F1 score, and Recall, achieving 97.63\%, 97.69\%, and 97.34\%, respectively. While other models perform well in certain aspects, none match the overall performance of the Ours model.
For instance, the ProtectAI model exhibits a slightly higher Precision at 99.59\%, but its Recall (79.83\%) and F1 score (88.62\%) are considerably lower than those of the Ours model, suggesting that while it excels in accurately identifying positive samples, it struggles with maintaining balance across the dataset. Similarly, the Hyperion model shows relatively balanced performance across all metrics, but its F1 score of 94.88\% is still below that of the Ours model.

Prompt Guard, Deepset, and Fmops demonstrate high recall but suffer from low accuracy and precision, indicating a tendency to misclassify benign samples as jailbreak attack samples. In contrast, WhyLabs LangKit achieves high accuracy and precision but at the cost of recall, suggesting it excels in classifying benign samples but struggles to accurately identify jailbreak attack samples. These performance characteristics highlight the limitations of these models, making them less suitable for practical deployment in real-world applications.

\begin{table}[ht]
\caption{Classification performance on Goal Hijacking Attack Scenarios.}
\centering
\setlength{\tabcolsep}{4pt} 
\renewcommand{\arraystretch}{1.4}  
\scriptsize   
\begin{tabular}{l c c c c}
\hline
\textbf{Method} & \textbf{Accuracy} $\uparrow$ & \textbf{Precision} $\uparrow$ & \textbf{F1} $\uparrow$  & \textbf{Recall} $\uparrow$ \\ \hline
\textit{ProtectAI}  & 94.25 & \textbf{99.79} & 93.95 & 88.76 \\ 
\textit{Hyperion}  & 90.68 & 94.53 & 90.33 & 86.48 \\ 
\textit{Prompt Guard} & 50.90 & 50.61 & 67.21 & \textbf{100} \\ 
\textit{Lakera AI} & 74.63 & 88.59 & 69.33 & 56.95 \\
\textit{Deepset} & 63.40 & 57.90 & 73.34 & \textbf{100} \\
\textit{Fmops} & 61.03 & 56.36& 72.09 & \textbf{100} \\
\textit{WhyLabs LangKit} & 68.14 & 97.53 & 54.35 & 37.67 \\
\hline
\textit{GenTel-Shield(Ours)} & \textbf{96.81} & 99.44 & \textbf{96.74} & 94.19 \\ \hline
\end{tabular}
\label{tab2}
\end{table}

\noindent\textbf{Goal Hijacking Attack Detection Results.}
Based on the experimental results in Table \ref{tab2}, our method outperforms existing approaches in several key metrics for detecting goal-hijacking attacks. While ProtectAI achieves a slightly higher Precision (99.79\%) compared to our method, its lower Recall of 88.76\% indicates a tendency to miss certain attack samples, potentially introducing security vulnerabilities. Prompt Guard, on the other hand, reaches a perfect Recall of 100\%, but its low Accuracy of 50.90\% reveals a significant issue with misclassifying benign samples as attacks. Similarly, both Deepset and Fmops achieve perfect Recall (100\%) but suffer from low Accuracy and Precision, further indicating an overclassification of benign samples as attacks.

In contrast, our method, GenTel-Shield, achieves the highest overall performance, with a best-in-class Accuracy of 96.81\% and F1 score of 96.74. Its high Precision (99.44\%) and Recall (94.19\%) suggest a strong balance between correctly detecting attack samples and minimizing false positives, making it more reliable for real-world application.

\begin{table}[h]
\caption{Classification Performance on Prompt Leaking Attack Scenarios.}
\centering
\setlength{\tabcolsep}{4pt} 
\renewcommand{\arraystretch}{1.4}  
\scriptsize   
\begin{tabular}{l c c c c}
\hline
\textbf{Method} & \textbf{Accuracy} $\uparrow$ & \textbf{Precision} $\uparrow$ & \textbf{F1} $\uparrow$  & \textbf{Recall} $\uparrow$ \\ \hline
\textit{ProtectAI} & 90.94 & \textbf{99.77} & 90.06 & 82.08 \\ 
\textit{Hyperion} & 90.85 & 95.01 & 90.41 & 86.23 \\ 
\textit{Prompt Guard}  & 50.28 & 50.14 & 66.79 & \textbf{100} \\ 
\textit{Lakera AI} & 96.04 & 93.11 & 96.17 & 99.43 \\
\textit{Deepset} & 61.79 & 57.08 & 71.34 & 95.09 \\
\textit{Fmops} & 58.77 & 55.07& 69.80 & 95.28 \\
\textit{WhyLabs LangKit} & \textbf{99.34} & 99.62 & \textbf{99.34} & 99.06 \\
\hline
\textit{GenTel-Shield(Ours)} & 7.92 & 99.42 & 97.89 & 96.42 \\ \hline
\end{tabular}
\label{tab3}
\end{table}

\noindent\textbf{Prompt Leaking
Attack Detection Results.}
Based on the experimental results in Table \ref{tab3}, our method shows excellent performance in detecting transient leak attacks. The WhyLabs LangKit slightly outperformed our method in terms of accuracy, precision, and F1 score, at 99.34\%, 99.62\%, and 99.34\%, respectively. This shows that WhyLabs LangKit has strong performance in leaking attack detection.
In comparison, our method strikes a balanced performance with a recall of 96.42\%, precision of 99.42\%, and F1 score of 97.89\%, maintaining high detection capability while reducing false positives. 
Additionally, our model significantly outperforms all other models, except for WhyLabs LangKit, across various performance metrics. This demonstrates that our model is also among the top performers in leak attack detection.

Overall, our model ranks among the best for detecting injection attacks, particularly in handling complex attack scenarios, showcasing exceptional classification capabilities. This highlights its strong potential for practical application in real-world settings.


\subsubsection{Separate Performance on Attack Samples} We evaluate the model's effectiveness on specific subsets of data. When the dataset consists solely of injection attack samples or specific scenes, we focus on accuracy as the key metric. This targeted evaluation allows us to understand how well the model performs on a particular type of data and understanding of the model's strengths and areas for improvement.




To evaluate the model's effectiveness in detecting jailbreak attacks and target hijacking attacks across various scenarios, we conducted experiments in \textit{Violation of Personal Rights}, \textit{Technology Misuse}, \textit{Social lmpact}, \textit{Serious Crime}, \textit{Infringement} and \textit{Ethical Violations} scenarios based on the security system outlined above. It is important to note that because \textit{Prompt Guard}, \textit{Deepset} and \textit{Fmops} tends to classify all samples as injection attacks, leading to a potential bias in comparison, we have chosen not to use them as a baseline in our experiment. Additionally, WhyLabs LangKit tends to misclassify injection attack samples as benign, resulting in poor performance on attack detection. As a result, we also exclude it from the experiment.

\begin{figure}[t]
  \includegraphics[width=0.5\textwidth]{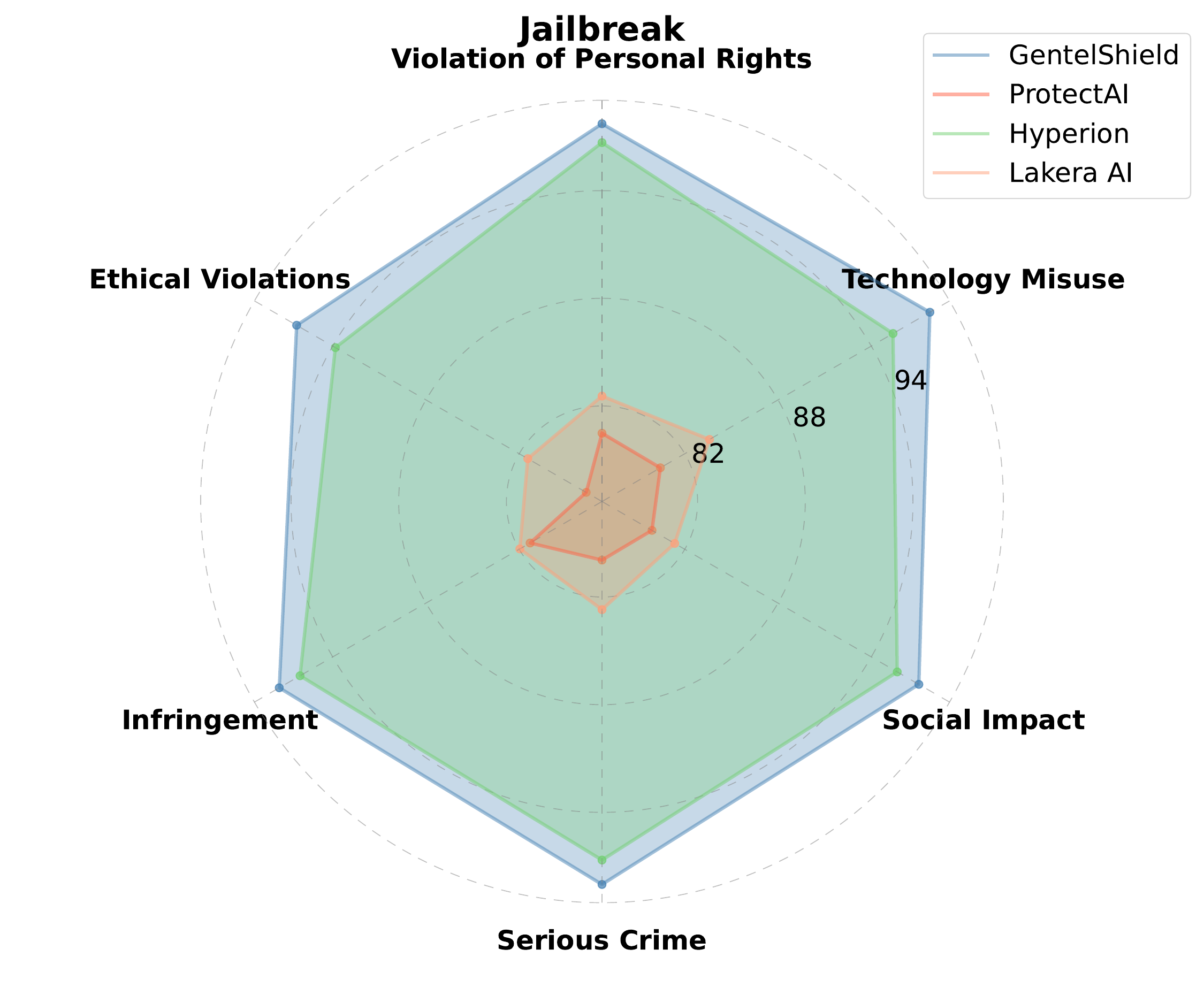}
  \caption{Classification Results for Jailbreak Attack Detection in different scenarios.}
  \label{fig:Jailbreak}
\end{figure}

As shown in Figure \ref{fig:Jailbreak}, in the jailbreak attack test in different scenarios,
 \textit{GenTel-Shield} emerges as the most balanced and robust model, suggesting its strong capability to handle a wide range of jailbreak attack scenarios. ProtectAI also performs well,  though it shows a slight dip in categories like "Ethical Violations," indicating room for improvement in handling more nuanced attack scenarios.
 Hyperion and Lakera AI show significant variability,  indicating a need for further refinement to ensure more consistent protection across different attack types.

\begin{figure}[t]
  \includegraphics[width=0.5\textwidth]{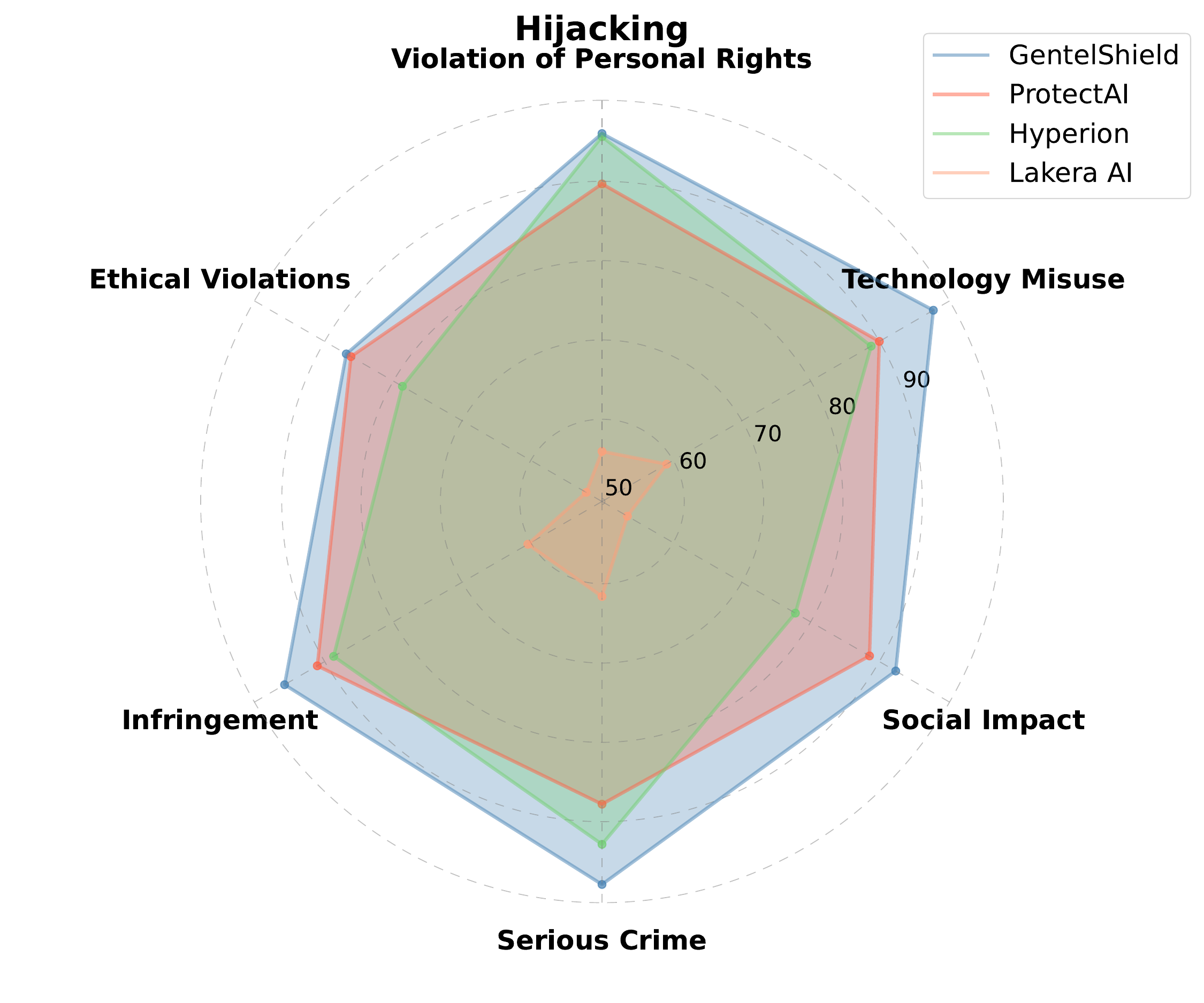}
  \caption{Classification Results for Hijacking Attack Detection in different scenarios.}
  \label{fig:Hijacking}
\end{figure}

\begin{figure*}[t]
  \includegraphics[width=\textwidth]{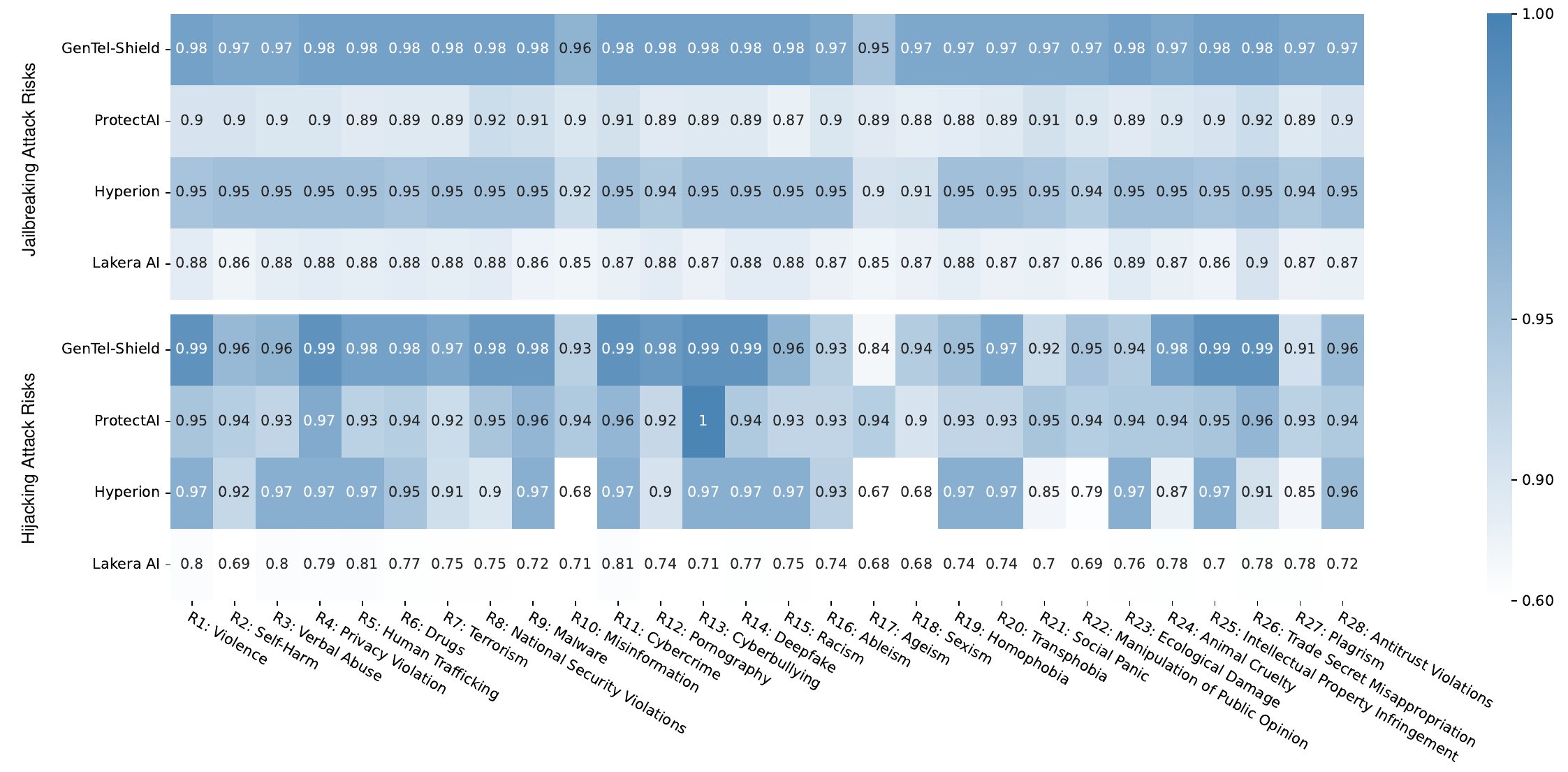}
  \caption{Identification accuracy under different risks. }
  \label{fig:heatmap}
\end{figure*}
Figure \ref{fig:Hijacking} show that classification results for Hijacking Attack Detection in different scenarios. \textit{GenTel-Shield} emerges as the most balanced and reliable model, consistently performing well across all categories. ProtectAI, while strong in certain areas like "Technology Misuse" and "Social Impact," has specific weaknesses that could affect its overall reliability. Hyperion shows particular strength in handling serious offenses but lacks consistency in other areas. Lakera AI, with its significant variability, may require further refinement to improve its performance in dealing with ethical and technological misuse challenges. Overall, \textit{GenTel-Shield} appears to be the most versatile model for addressing a wide range of hijacking attack scenarios.

In order to explore the performance of different models in various subdivision scenarios, we conducted a series of experiments.The experimental results presented in the Figure. \ref{fig:heatmap} highlight the performance of five different models—GenTel-Shield, ProtectAI, Hyperion, Lakera AI, and WhyLabs LangKit—across various risk categories, including Jailbreaking Attack Risks and Hijacking Attack Risks.

GenTel-Shield consistently outperformed other models in most categories, achieving high accuracy across several risk scenarios, including Violence (R1), Privacy Violation (R4), and Cybercrime (R11). It maintained an overall performance rate near 0.99 in these categories. Similarly, ProtectAI showed strong performance in specific domains like Misinformation (R10) and Terrorism (R7), but slightly lagged behind in areas like Malware (R9).

Lakera AI and Hyperion demonstrated moderate consistency, with Hyperion maintaining higher performance levels in areas such as Racism (R15) and Deepfake (R14). Lakera AI struggled in some categories, particularly Ageism (R17) and Homophobia (R19), where performance metrics dropped below 0.75.

In summary, the experiments demonstrate that our \textit{GenTel-Shield} model effectively detects injection attacks across various security scenarios, highlighting its strong practical value and versatility.

\subsection{Comparison Results on Different Datasets}

To demonstrate the generalization ability of the model, we conducted experiments on multiple datasets. Table \ref{tab5} presents the classification performance of various models on the Jailbreak-LLM and Deita datasets. Jailbreak-LLM consists of 1,405 attack samples, while Deita contains 9,500 normal instructions. We merged these two datasets for the experiments.

The results show that our model consistently outperforms all others across key metrics, particularly in Accuracy and F1 score, achieving 97.26\% and 90.35\%, respectively. While some models excel in specific areas, their overall performance remains subpar. For example, the Deepset model demonstrates a slightly higher Precision of 99.72\%; however, its Accuracy (50.99\%) and F1 score (34.39\%) are significantly lower than those of our model. This suggests that while Deepset is highly effective at identifying positive samples, it struggles with overall dataset balance.

Similarly, the ProtectAI model shows relatively balanced performance across all metrics, yet its F1 score of 86.67\%, Precision of 85.84\%, and Accuracy of 96.60\% still fall short of our model's results. Notably, the Hyperion model exhibits relatively poor performance, particularly in terms of Accuracy and Precision, highlighting its potential limitations when applied to jailbreak attack scenarios.

\begin{table}[h]
\caption{Classification performance on Comparison Results on Other Datasets.}\label{tab5}
\centering
\setlength{\tabcolsep}{4pt} 
\renewcommand{\arraystretch}{1.4}  
\scriptsize   
\begin{tabular}{l c c c c}
\hline
\textbf{Method} & \textbf{Accuracy} $\uparrow$ & \textbf{Precision} $\uparrow$ & \textbf{F1} $\uparrow$  & \textbf{Recall} $\uparrow$ \\ \hline
\textit{ProtectAI}  & 96.60 & 85.84 & 86.67 & \textbf{87.52} \\ 
\textit{Hyperion}  & 55.12 & 49.90 & 22.27 & 14.33 \\ 
\textit{Prompt Guard} & 20.50 & 91.32 & 22.84 & 13.05 \\ 
\textit{Lakera AI} &93.21 & 82.28 & 75.75 &70.19 \\
\textit{Deepset} &50.99 & \textbf{99.72} & 34.39 & 20.78 \\
\textit{Fmops} & 47.39 & 99.29 & 32.72 & 19.59 \\
\textit{WhyLabs LangKit} & 94.54 & 69.89 & 76.75 & 85.10 \\
\hline
\textit{GenTel-Shield(Ours)} & \textbf{97.26} & 99.57 & \textbf{90.35} & 82.68 \\ \hline
\end{tabular}
\label{tab5}
\end{table}


\subsection{Comparison Results on Different Language}

To assess the model's generalization performance across different language datasets, we used Google Translate to translate the datasets for jailbreak attacks, target hijacking attacks, and prompt leakage scenarios into Chinese, Japanese, French, Spanish, and German. The translated datasets were then used for testing.

\begin{figure}[h]
  \includegraphics[width=0.5\textwidth]{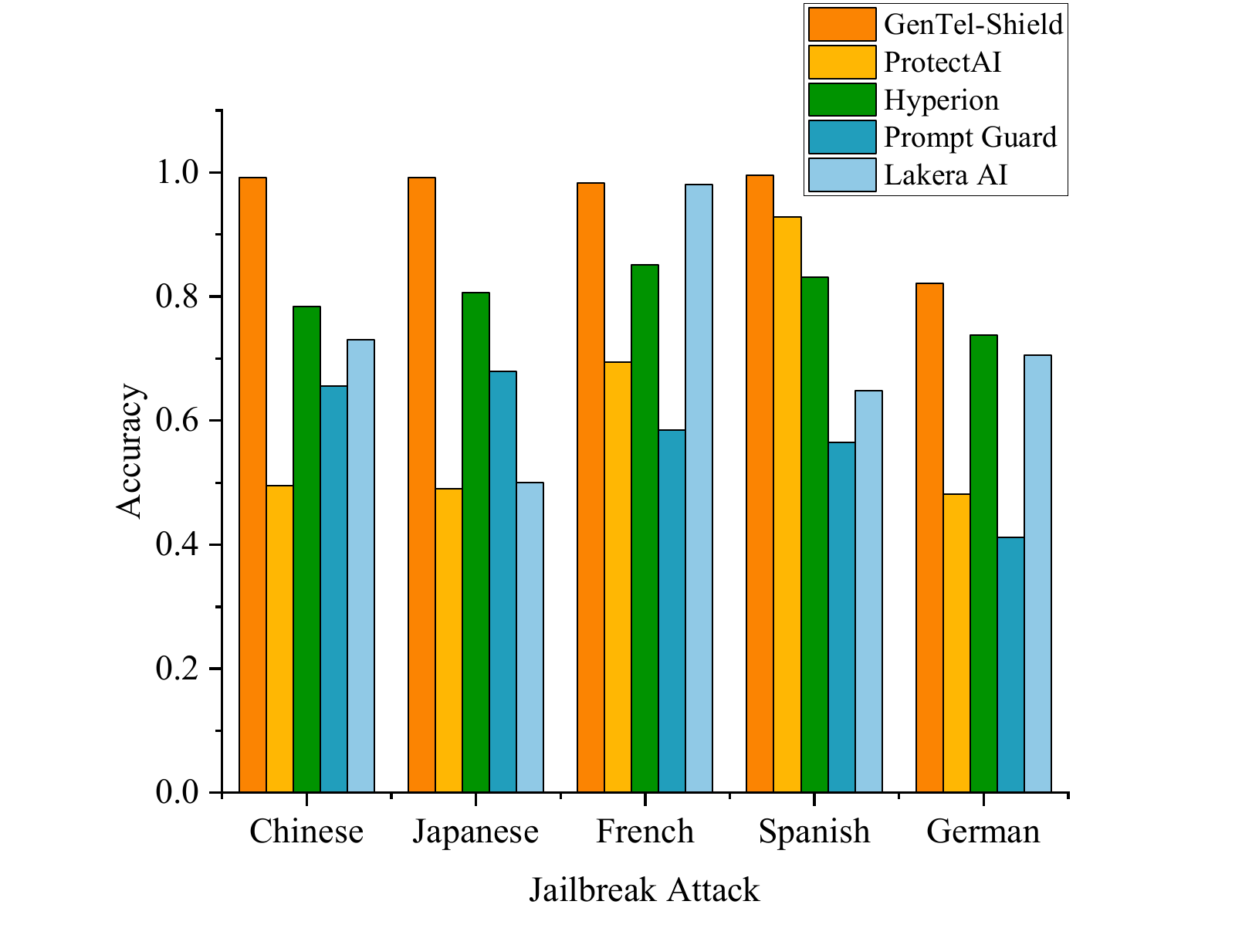}
  \caption{Jailbreak experiment on different language.}
  \label{language:Jailbreak}
\end{figure}

Figure \ref{language:Jailbreak} illustrates the defensive performance of various injection attack detection models in the jailbreak attack scenario. The results highlight the differences in generalization ability across languages for different models. For instance, while Lakera AI performs well in French, its effectiveness is lower in other languages. In contrast, our model, GenTel-Shield, demonstrates top-tier performance across all languages, with jailbreak detection accuracy in Chinese, Japanese, French, and Spanish approaching nearly 100\%.

\begin{figure}[h]
  \includegraphics[width=0.5\textwidth]{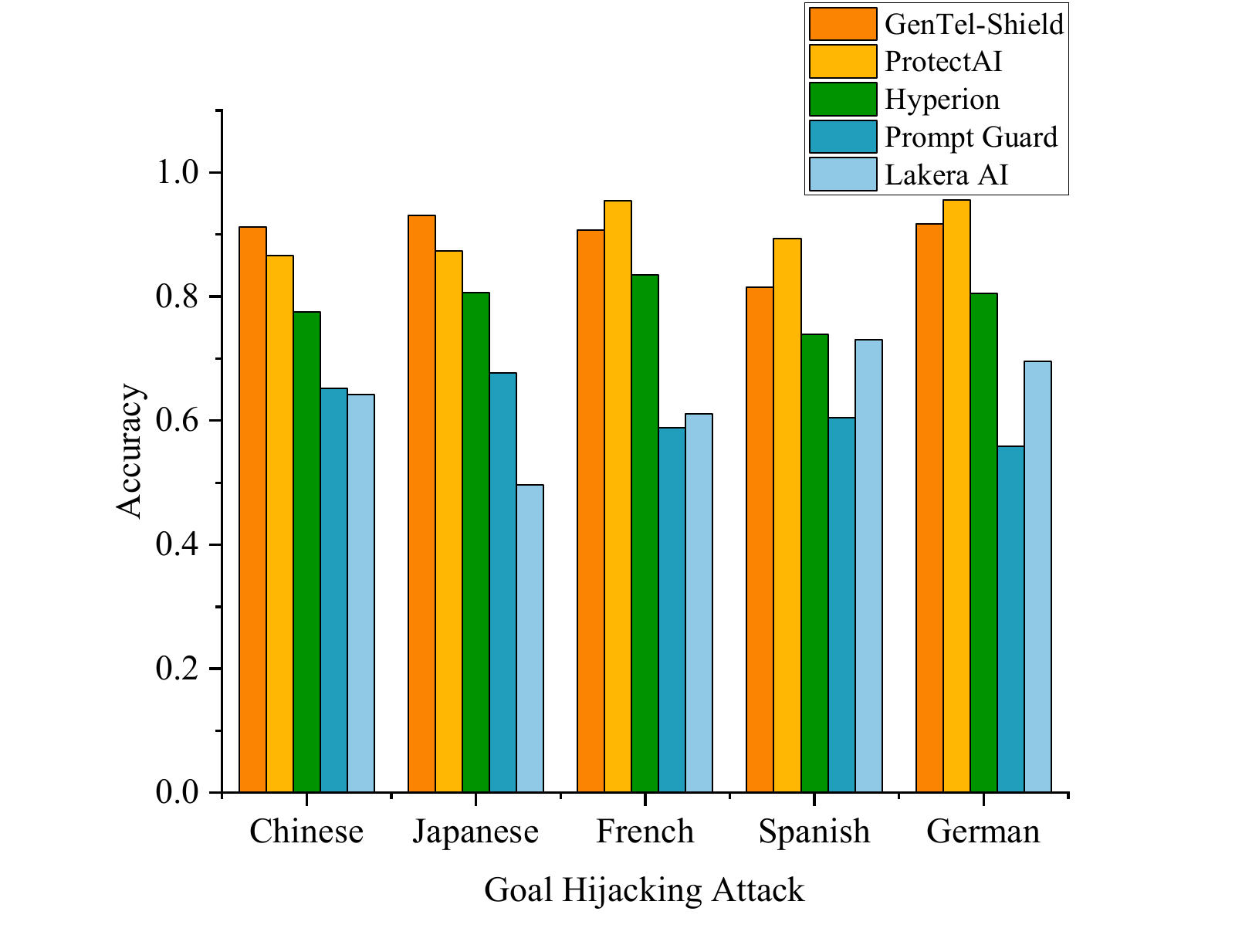}
  \caption{Goal hijacking experiment on different language.}
  \label{language:Hijacking}
\end{figure}

Figure \ref{language:Hijacking} highlights the variations in detection model performance for target hijacking attacks across different languages. GentelShield demonstrates strong defense capabilities, achieving the highest accuracy in Chinese and Japanese. ProtectAI performs exceptionally well in French, Spanish, and German, even surpassing GentelShield in these languages. Hyperion exhibits a more balanced performance. Lakera AI performs well in Spanish and German but struggles in other languages, especially Japanese. Overall, the performance of GentelShield is similar to that of ProtectAI, with outstanding results in Chinese and Japanese.

\begin{figure}[t]
  \includegraphics[width=0.5\textwidth]{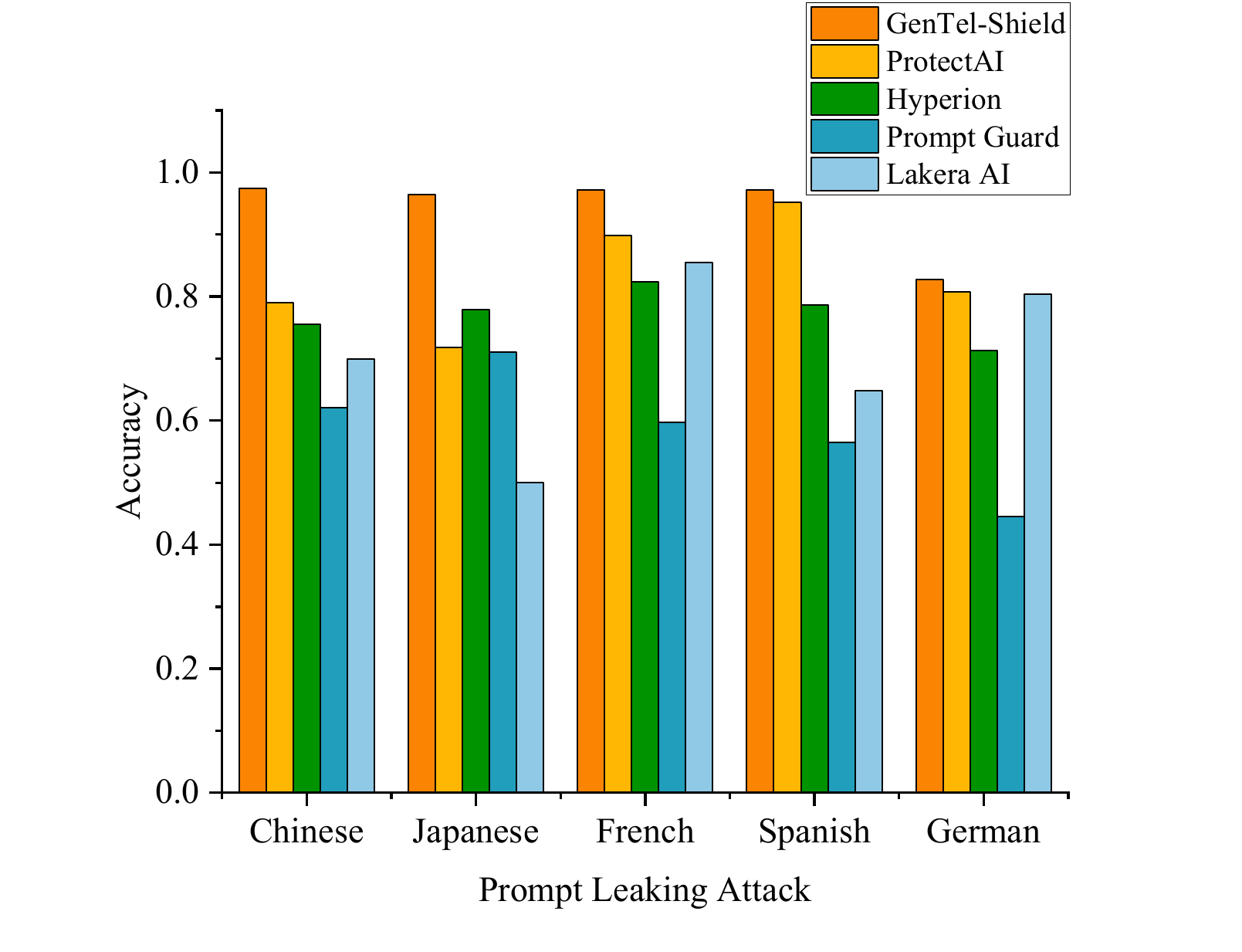}
  \caption{Prompt leaking experiment on different language.}
  \label{language:Leaking}
\end{figure}

Based on the experimental results Figure \ref{language:Leaking} provided, GentelShield stands out with the highest accuracy rates across four languages—Chinese, Japanese, French, and Spanish—with its peak performance in Chinese at an impressive 97.36\%. This indicates that GentelShield has a high level of accuracy and stability when handling multilingual tasks.
The ProtectAI model follows closely behind, demonstrating high accuracy rates in all tested languages except for French, with its best performance in Spanish, where it achieved an accuracy rate of 95.19\%, just behind GentelShield. This suggests that ProtectAI is a strong contender and may have specific advantages in certain languages.
The MetaPromptGuard model shows significant fluctuations in accuracy rates, ranging from 44.53\% in German to 71.04\% in Japanese, indicating inconsistent performance across different languages.
Hyperion demonstrates a more balanced performance, with an accuracy rate of 77.92\% in Japanese, second only to GentelShield.

\subsection{Comparison Experiment with Different Input Token Lengths}
To examine the impact of varying input token lengths on the model's performance, we conducted tests using token lengths of 128, 256, 384, and 512, respectively.
Since WhyLabs LangKit and Lakera AI are accessed via API calls and do not allow for modification of the token length, these two models are excluded from the following experiments.

\begin{figure}[t]
  \includegraphics[width=0.5\textwidth]{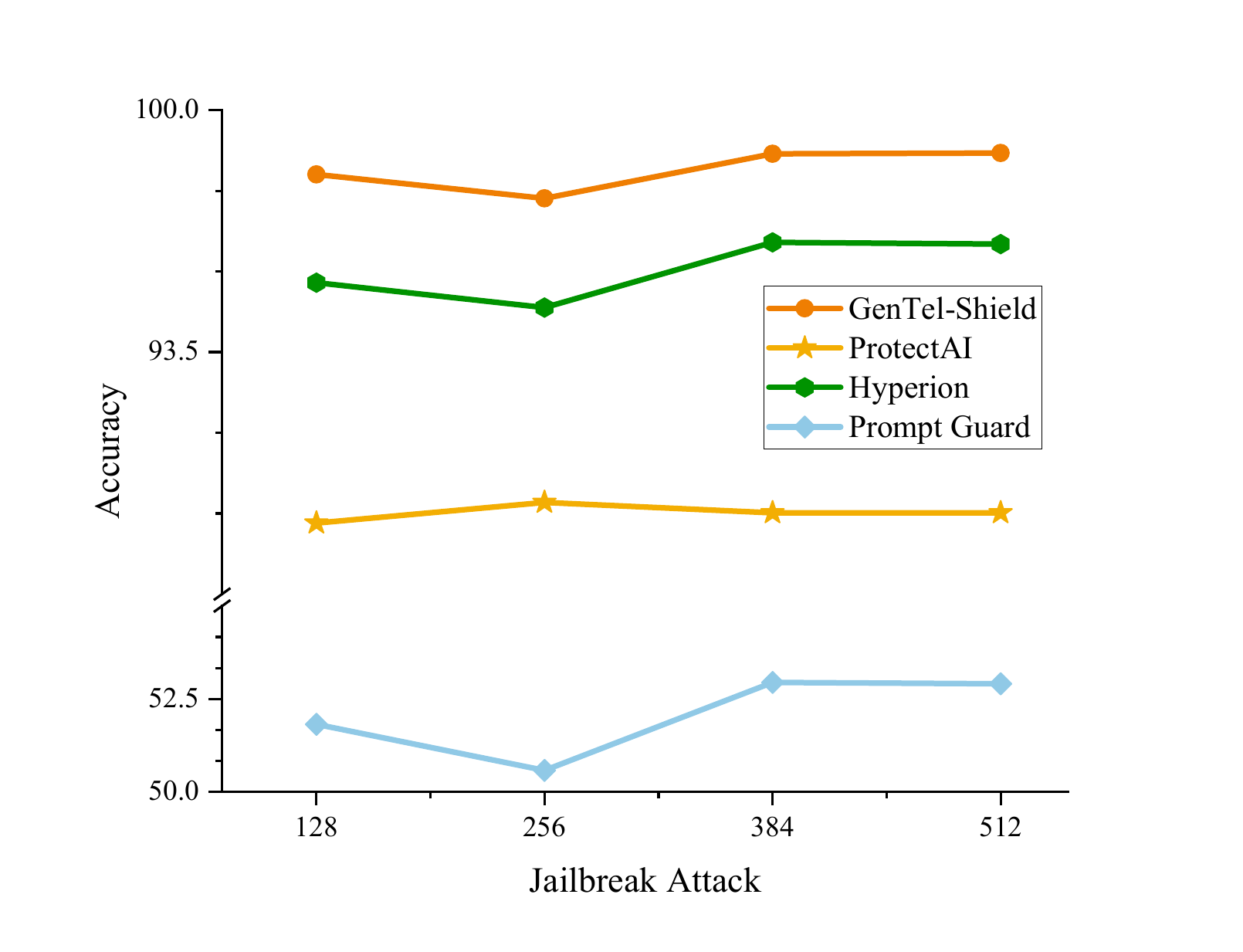}
  \caption{Jailbreak experiment on Token length.}
  \label{length:Jailbreak}
\end{figure}

As shown in Figure \ref{length:Jailbreak}, on the jailbreak attack test dataset, our \textit{Gentel-Shield} consistently outperforms other models, regardless of the token length setting. An interesting observation is that when the input token length increases from 128 to 256, the performance of some models decreases to a certain extent. This may be because a token length of 128 captures only partial jailbreak attack templates or security issues, leading to higher variability in the model's predictions. However, when the token length is increased to 384, the performance of most models improves significantly, as 384 tokens can cover most of the jailbreak templates and related issues, allowing the models to process a more complete prompt. Beyond 384 tokens, further increasing the token length to 512 results in minimal changes in performance.

\begin{figure}[h]
  \includegraphics[width=0.5\textwidth]{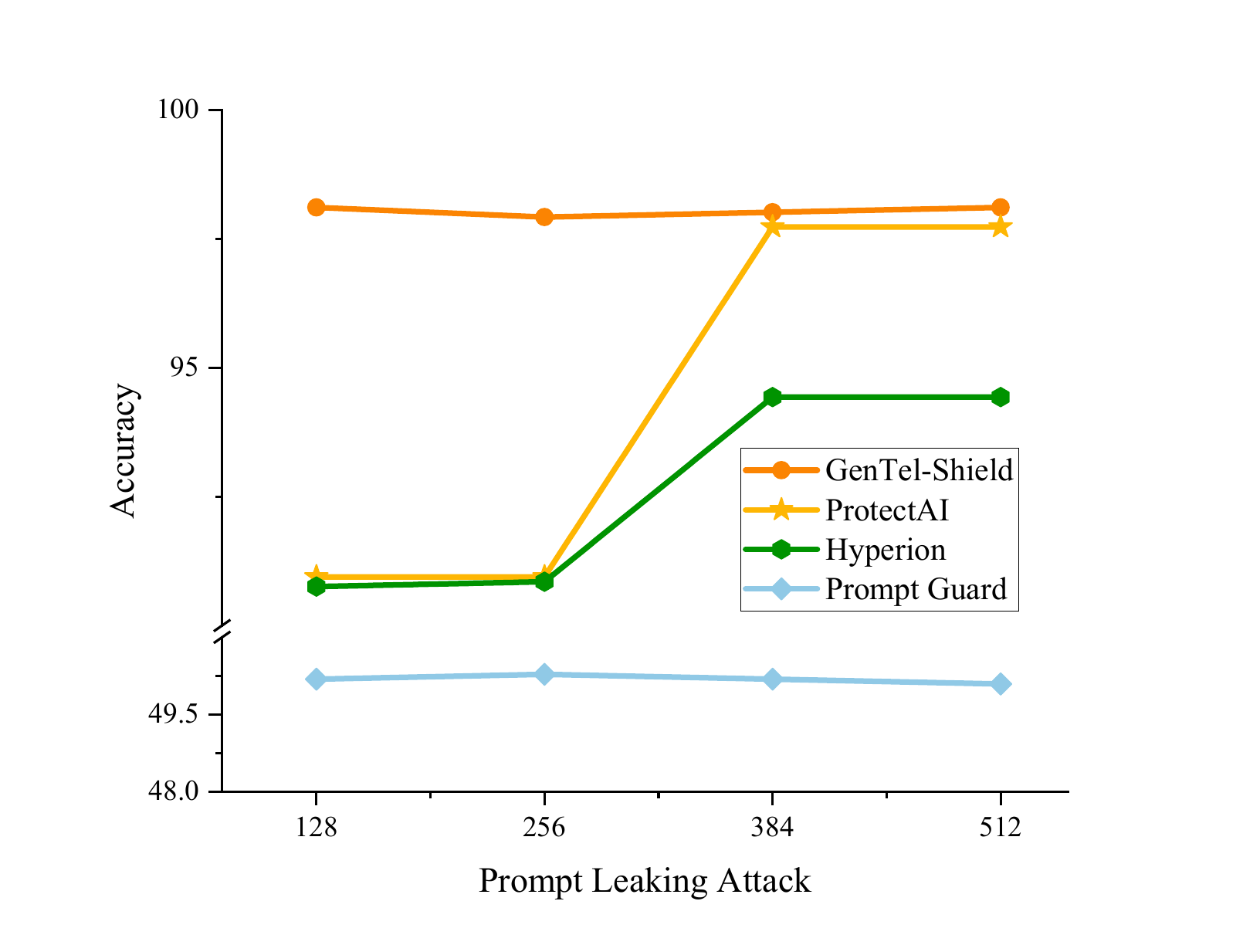}
  \caption{Prompt leaking experiment on Token length.}
  \label{length:Leaking}
\end{figure}

As shown in Figure \ref{length:Leaking}, this phenomenon is more pronounced in the dataset with prompt word leakage. Since prompt leakage attacks are more diverse than jailbreak attacks, the additional prompts do not consistently appear at the beginning or end of the input. This variability makes it harder for models to accurately detect prompt leakage, further highlighting the impact of token length on performance.

In summary, the input token length limit affects model performance to some extent, but once the token length reaches a certain threshold, its impact diminishes, and the performance of each model tends to stabilize.

\section{Conclusion}

In this work, we present a comprehensive framework for training models to defend against prompt injection attacks, called \textit{GenTel-Shield}, and introduce \textit{Gentel-Bench} for assessing defending methods' effectiveness. Our study demonstrates \textit{GenTel-Shield}'s robust capability in identifying attack and benign application cues. Comprehensive evaluations across three distinct attack scenarios underscore the versatility and effectiveness of \textit{GenTel-Shield}. We hope our proposed LLM security risk benchmark and defense shield will inspire further research into diversifying LLM defending strategies.

\bibliography{custom}

\appendix



\end{document}